\documentclass[cameraready]{Interspeech}

\title{VIP-MINGLE: A Corpus for Videoconference and In-Person Multimodal Interaction in Group Language Engagement}

\author[affiliation={1}, orcid=0000-0001-6745-4435, correspondingauthor]{Andrew}{Chang}
\author[affiliation={1}, orcid=0009-0009-6423-4002, equalcontribution]{Abhinay K}{Bodi}
\author[affiliation={1}, orcid=0009-0003-8680-1502, equalcontribution]{Wenxin}{Deng}
\author[affiliation={1}, orcid=0009-0001-2749-5440, equalcontribution]{Junrui}{Huang}
\author[affiliation={1}, orcid=0009-0003-3356-7628, equalcontribution]{Venu G}{Kadamba}
\author[affiliation={1}, orcid=0009-0001-9166-6494, equalcontribution]{Sumanth B H}{Karanam}
\author[affiliation={1}, orcid=0009-0001-0497-7316, equalcontribution]{Dhiwahar A}{Kennady}
\author[affiliation={1}, orcid=0000-0003-0184-163X]{David}{Poeppel}
\author[affiliation={1}, orcid=0000-0002-3934-9723]{Dustin}{Freeman}

\address{
    $^1$ New York University, USA
}

\email{c.andrew123@gmail.com, ab11958@nyu.edu, wd2261@nyu.edu, jh8186@nyu.edu, vk2636@nyu.edu, sh8111@nyu.edu, dk5025@nyu.edu, dp101@nyu.edu, dustin.freeman@gmail.com}


\usepackage{comment}

\begin{document}

\maketitle

\begin{abstract}
    Group conversations are a fundamental yet complex form of social interaction central to human cognition and telecommunication technology. While understanding and facilitating these interactions has been a long-standing goal, findings are often isolated within specific in-person or videoconferencing settings due to a scarcity of datasets that bridge the two. We introduce VIP-MINGLE, a multimodal dataset comprising 59 hours of recordings (32 groups, 105 participants), featuring paired within-subject sessions in both settings. The dataset includes raw audio/video, psychometric data, processed multimodal features (e.g., diarized speech, facial expressions, transcriptions), and time-resolved human annotations. Our analysis reveals significant behavioral distribution shifts across multiple modalities between settings, reinforcing the need for a cross-setting corpus. VIP-MINGLE serves as a critical resource for developing robust models of group conversations across settings.
\end{abstract}

\section{Introduction \& Relevant Works}

The rapid digitization of the workplace has fundamentally transformed human communication. Videoconferencing platforms have evolved from occasional alternatives into a primary medium for professional collaboration, education, and social interaction. As this hybrid reality stabilizes, there is a need for systems that can understand, analyze, and support group interactions across both physical and remote settings.

Despite substantial progress in relevant technologies, current research faces a critical domain shift challenge. Historically, models have been trained and evaluated on high-fidelity, in-person conversational corpora—most notably the AMI and ICSI Meeting Corpora—where social signals such as gaze direction, spatial audio cues, and fluid backchanneling are naturally preserved \cite{kraaij2005ami, janin2003icsi}. However, these assumptions often break down in videoconferencing settings. The discrepancy is not merely the result of technical artifacts such as compression or transmission delay; rather, it reflects a fundamental restructuring of human interaction. Prior work has shown that videoconferencing fundamentally reshapes turn-taking, attention, and non-verbal behavior in ways that extend beyond signal-level distortions \cite{fauville2021zoom, boland2022zoom, fauville2023video}. Therefore, videoconferencing should be viewed as a qualitatively distinct form of in-person conversation, rather than simply a degraded approximation.

Recognizing these shifts, recent efforts have introduced datasets capturing naturalistic videoconferencing interactions. For instance, the CANDOR corpus contains thousands of dyadic conversations, and the RoomReader corpus provides 30 group sessions with multimodal features \cite{reece2023candor, reverdy-etal-2022-roomreader}. However, these valuable resources are isolated within the remote domain. To bridge this gap, some recent works have attempted cross-setting comparisons. For example, Tian et al. \cite{tian2024corpus} identified divergent turn-taking dynamics by comparing existing recordings from bilibili.com; however, relying on naturally occurring, unpaired data introduces confounding factors—such as task or protocol variations—that make it difficult to isolate the medium's true effect. Other studies have used tightly controlled experimental designs to contrast the two settings \cite{boland2022zoom, balters2023virtual}, but they focused primarily on the temporal aspects of speech and did not provide publicly accessible recordings. Consequently, disentangling the effects of the communication medium from speaker-specific traits, task-level variations or dataset-level artifacts remains a complex methodological challenge.

To address this gap, we introduce the \textbf{VIP-MINGLE} (Videoconference and In-Person Multimodal INteraction in Group Language Engagement) corpus. Building on previous modeling works that have demonstrated the critical role of multimodal features in capturing conversation and group dynamics \cite{chang2025icassp, chang2025interspeech, subburaj2020multimodal, stewart2021multimodal, kendrick2023turn}, VIP-MINGLE employs a controlled, within-subjects design in which the same groups of participants engage in identical collaborative tasks across both in-person and videoconferencing settings. This controlled design enables direct, participant-level comparisons of conversational behavior, providing a unique testbed for rigorously quantifying domain shift and developing models that can adapt robustly across physical and remote group interactions.

\begin{figure*}[ht]
  \centering
  \includegraphics[width=0.9\linewidth]{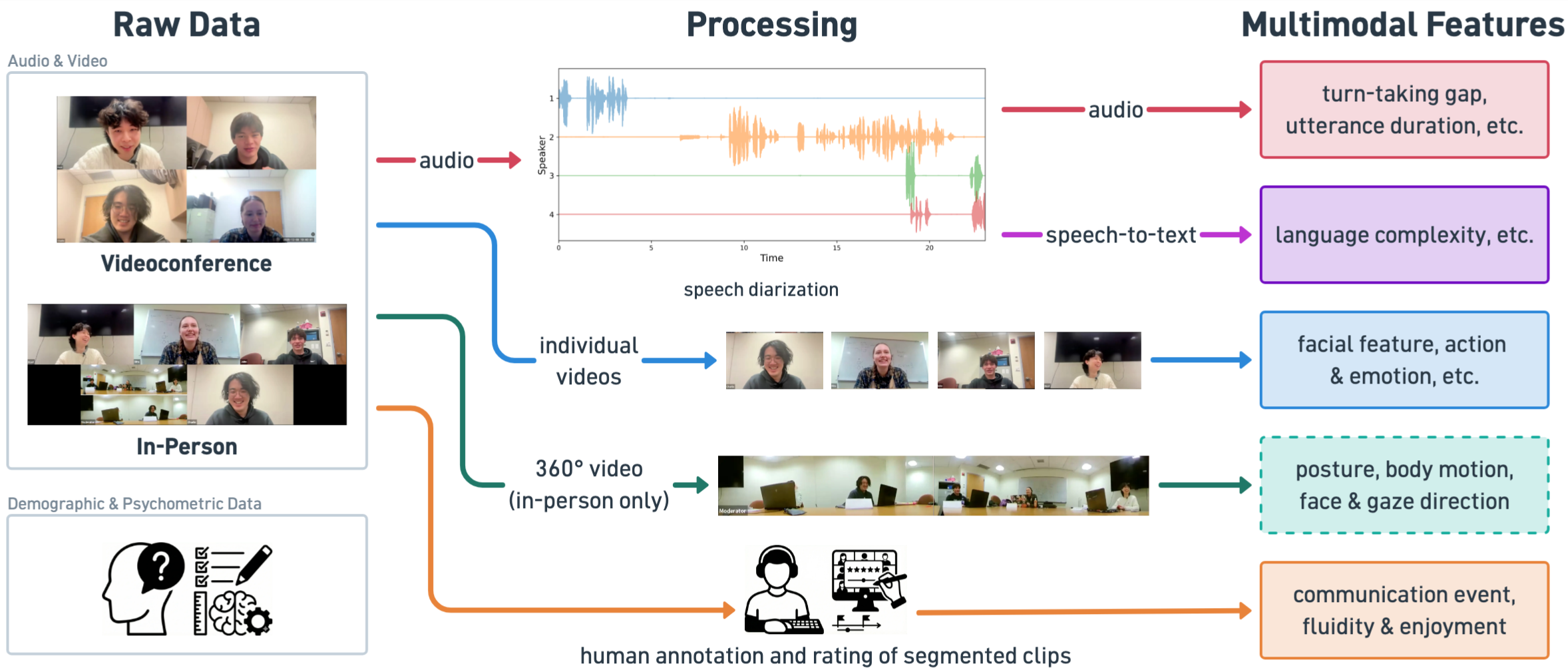}
  \caption{Schematic of the VIP-MINGLE data, signal processing pipeline and multimodal features across in-person and videoconference settings. Solid boxes: provided raw recordings and processed features; Dashed box: derivable features.}
  \label{fig:fig1}
\end{figure*}

\section{VIP-MINGLE Corpus}

Unique to VIP-MINGLE is its within-subject design. Every group participates in two distinct communication settings: in-person and videoconference (remote), allowing for direct intra-group comparison across modalities. It includes both group-level and participant-level multimodal features (e.g., raw audio/video, diarized speech, transcripts, facial expressions), supplemented by standardized participant-level psychometric baselines for personality and mood \cite{gosling2003tipi, watson1988panas} a few days prior to the experimental sessions. 

In total, the corpus comprises $\approx$59 hours of recordings, an average session duration of 21 minutes, 105 participants across 32 groups, and 7,077 annotated segments. The scripts and dataset are available at DOI: 10.5281/zenodo.20670131.

\subsection{Participants, Procedure and Task}
Participants (age: 17–28; gender: 29 men, 69 women, 7 other) were recruited from New York University. Participants received course credit for their time, and they provided written informed consent regarding data collection and sharing (IRB approved).

Recruited participants were assigned to groups of 2--4. We utilized a within-subjects design where each group completed both an in-person and a videoconference session, with the order counterbalanced to control for order effects. The total procedure lasted under two hours. The primary task was a “Family Feud”-style game, as used in prior work, to elicit unconstrained turn-taking and collaborative problem-solving \cite{reverdy-etal-2022-roomreader, koutsombogera2018modeling}. A researcher served as the initial game host. During the game, participants took turns acting as the "host" of that round, asking trivia questions drawn from a randomized list while others answered freely. All conversations were conducted in English. To maintain privacy, aliases were used throughout the sessions.

\subsection{Recording Setup}

Both in-person and videoconferencing sessions were recorded via Zoom (individual video/audio tracks, Active Video Only (AVO), and Gallery Video Only (GVO)) using personal laptops and external USB lavalier microphones (MAONO AU-UL10) clipped below the chin. Videoconferencing participants were placed in separate, isolated rooms. In contrast, in-person participants sat around a shared conference table with $\sim$1--3 meters of separation. To minimize interference during in-person sessions, laptop screens were dimmed, speakers were muted, and a 360° camera (j5create JVCU360) was centered on the table. We ensured that none of the devices occluded visual contact among the participants. While the target-to-interfering speaker distance ratio was well below the standard 3:1 rule of thumb, microphone gain was manually reduced during recording to minimize audio crosstalk and subsequently restored during signal processing.

\subsection{Signal Processing and Feature Extraction}

To explore how different communication settings influence behaviors, we utilized pretrained models for feature extraction across each modality. Our processing pipeline transforms raw audio and video into temporally aligned, participant-level representations spanning the audio, visual, and textual domains (Figure \ref{fig:fig1}). Recognizing the multitude of derivable features, we selected well-established models to extract a representative subset across several modalities. This provides a practical baseline for cross-setting comparison within this corpus, which also enables comparability with previous corpora and studies (e.g., \cite{reece2023candor, reverdy-etal-2022-roomreader, chang2025icassp}). Both raw and processed data are included in the VIP-MINGLE corpus, providing researchers the flexibility to apply their preferred or future state-of-the-art models for custom feature extraction and downstream analyses.

\begin{figure}[!ht]
  \centering
  \includegraphics[width=1\linewidth]{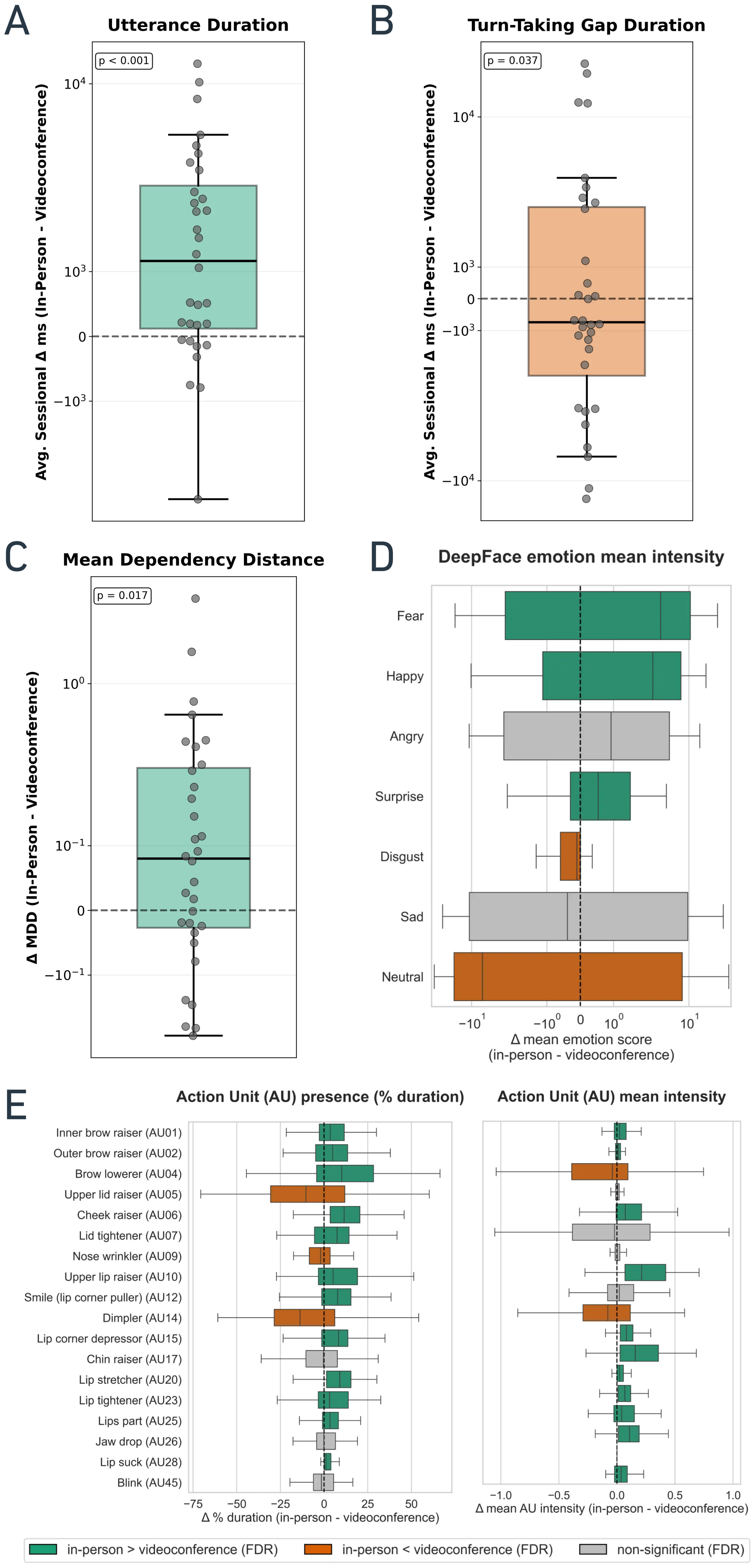}
  \caption{Pairwise behavioral feature comparisons between settings: (A, B) speech temporal features, (C) language (syntactic) complexity, and (D, E) facial expressions (note: OpenFace does not support intensity data for AU28). The results show a broad spectrum of substantial behavioral shifts. Specifically, videoconferencing elicits shorter, less complex speech with longer pauses. Facially, it is associated with generally less salient expressions, yet exhibits an increase in specific localized ones.}
  \label{fig:compare_features}
\end{figure}

\subsubsection{Audio \& Speaker Diarization}

Audio tracks were intrinsically isolated by room during videoconferencing sessions. For in-person sessions, we isolated individual participant audio by applying the pyannote/speaker-diarization pipeline \cite{Bredin23, Plaquet23} to extract time-stamped speaker segments. Specifically, the recordings were first downsampled to 16 kHz mono WAV format to optimize diarization performance. The diarization pipeline identified the optimal microphone for each participant based on proximity indicators, including high-frequency content, total speaking time, and spectral characteristics. We isolated the primary speaker's signal from their respective dedicated microphone. The pipeline then extracted the time-aligned audio for each participant—preserving exact conversational timing by padding non-speaking intervals with silence—and encoded the final signals as mono, 32 kHz audio-only MP4 files to ensure high audio fidelity. These signals were later remixed and realigned with videos for human annotation.

\subsubsection{Speech Transcription}
We first standardized audio formats to mono WAV and sampling rates at 16 kHZ, and then used the Whisper (large) model \cite{radford2022whisper} to generate noise-robust transcripts with word-level or segment-level timestamps. It resulted in JSON or text outputs into a standardized representation. We align each transcript segment with diarized speaker labels, thereby producing speaker-attributed, time-stamped text. Text processing scripts further clean and organize transcripts (e.g., normalizing punctuation, handling overlapping speech) to facilitate downstream analysis of discourse structure and linguistic style.

\subsubsection{Video \& Facial Feature Extraction}

Individual video streams were extracted from Zoom GVO recordings using an automated pipeline with manual quality verification. We prioritized GVO tiles over 360° room cameras because they provide higher effective facial resolution and more stable, near-frontal views—factors essential for reliable feature and expression analysis. We extracted two complementary feature sets to capture both geometric behavior and high-level emotion:

\textit{OpenFace}: We utilized OpenFace \cite{openface} to estimate 3D head pose, eye-gaze direction, and action unit (AU) intensities and presences. Analysis was conducted at the native frame rate ($\approx$ 25--30 fps) to preserve temporal dynamics. To maintain signal integrity, infrequent frames with failed detections (e.g., due to occlusion) were flagged as missing rather than imputed.

\textit{DeepFace}: We used DeepFace \cite{serengil2026boosted} to extract probabilities time series of each emotion (e.g., happy, sad). These data were synchronized with OpenFace features via a common frame index. The final output is a unified, frame-level feature matrix for each participant, integrating interpretable geometric markers with deep affective representations.

\subsubsection{360° Video Recordings}
To capture a unified spatial reference frame, we supplemented in-person sessions with 360° video recordings, allowing for the extraction of gaze, head, and body orientation and movement data, a level of synchronized spatial detail unavailable through Zoom’s GVO.

\section{Comparing In-Person and Videoconference Multimodal Features}

\begin{figure*}[t]
  \centering
  \includegraphics[width=\linewidth]{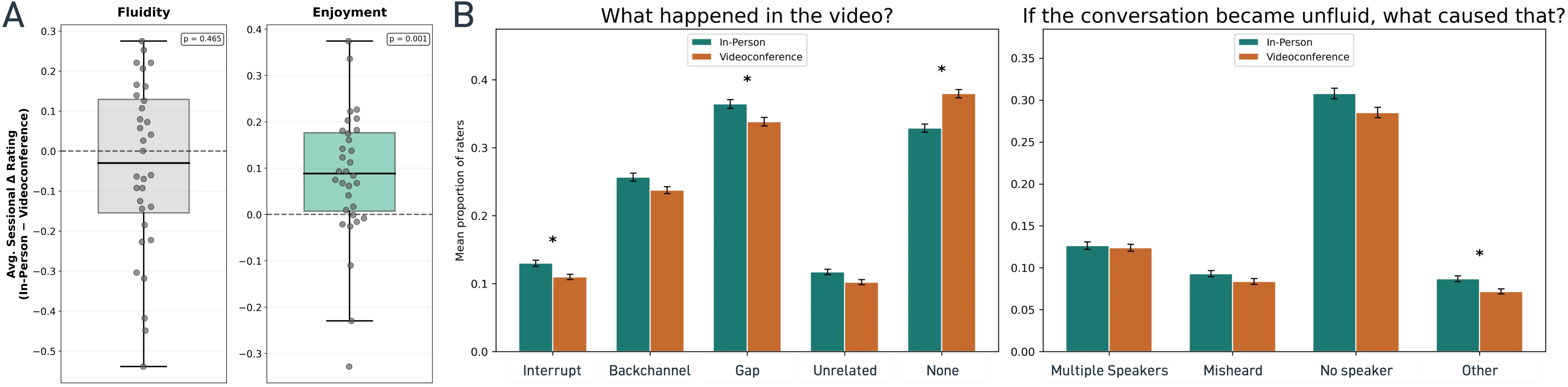}
  \caption{Human-rated conversational qualities and annotated events. Ratings and annotations were performed at the segment level and averaged per session for statistical analysis. (A) In-person sessions scored significantly higher in enjoyment than videoconferencing, despite no significant difference in conversational fluidity. (B) In-person sessions exhibited a higher frequency of interruptions and gaps, suggesting that such conversational "turbulence" may be positively associated with overall enjoyment.}
  \label{fig:annotation_results}
\end{figure*}

To investigate whether videoconferencing simply degrades in-person communication or features a substantially different behavioral pattern, we conducted pairwise comparisons of multimodal features between settings. Even within a representative subset of all derivable features, our exploratory results show substantial behavioral changes rather than mere scaling differences. These highlight the complex domain gap that must be addressed when modeling conversations across settings.

\subsection{Speech Temporal Features}
To quantify conversational dynamics, we utilized the Heldner–Edlund model \cite{heldner2010pauses} to extract speaker turn boundaries, calculating turn-taking gap duration and utterance duration. Data were log-transformed to mitigate right-skewness, and the linear mixed-effects model analyses were used to control for within-group dependency. Results (Figure \ref{fig:compare_features}AB) indicate that videoconference sessions exhibited significantly longer turn-taking gaps ($\beta =0.113, SE=0.054, p=.037$), consistent with prior findings \cite{boland2022zoom, tian2024corpus}. Also, the videoconference sessions had a shorter utterance duration ($\beta=-0.094, SE=0.018, p<.001$). In short, the conversational pace in the videoconference setting tends to be slower and more fragmented.

\subsection{Language Complexity}

To examine whether in-person and videoconference interactions differed in conversational language complexity, we computed session-level complexity metrics from the per-speaker transcripts. For each session, we calculated each metric separately for every participant's transcript and then averaged across participants to obtain a single session-level value per group. We calculated four metrics: (1) Mean Dependency Distance (MDD), a measure of syntactic complexity, computed using the TextDescriptives package with spaCy's English dependency parser \cite{Hansen2023}; (2) measure of textual lexical diversity \cite{LexicalRichness}; (3) mean token surprisal; and (4) perplexity, both estimated using a pretrained language model via the minicons \cite{misra2022minicons}. Wilcoxon signed-rank test showed that MDD was significantly higher in the in-person than in the videoconference session ($p=.017$; Figure \ref{fig:compare_features}C), indicating greater syntactic complexity in in-person communication. No significant differences were found for other metrics ($p\geq.410$). These results suggest that while the semantic metrics may be approximate, participants tend to use simpler syntactic structures during videoconferencing.

\subsection{Facial Expressions}
We performed Wilcoxon signed-rank tests on individual participant's OpenFace AU metrics (excluding researchers' data), including the percentage of presence duration and mean intensity. While the majority of AUs displayed heightened activation in in-person sessions, a few showed significant gains in videoconference sessions. The same tests on DeepFace's emotional intensity showed that the happy, surprise and fear expressions were more salient in in-person setting while the disgust and neutral expressions were more salient in videoconference sessions (Figure \ref{fig:compare_features}DE). Results show different settings don't just scale facial expressions; they elicit qualitatively distinct patterns.

\section{Rated Conversational Interactions \& Annotated Events}

Videos were segmented into 10-second clips and annotated by 192 qualified annotators on Qualtrics (after excluding those with low reliability \cite{chang2025icassp}), with each annotator viewing 120 clips. Participants were crowdsourced from New York University for course credits (aged 18--25; 121 women, 44 men, 27 other) and were instructed to complete the annotations in a quiet space at their own pace. This study was approved by the IRB.

Analysis of 7,077 rated clips (5-point Likert scale) revealed that conversations in in-person setting were rated significantly higher in enjoyment than videoconferencing (Wilcoxon signed-rank tests at the sessional level: $p = .001$), while no significant difference was found regarding conversational fluidity ($p = .465$) (Figure \ref{fig:annotation_results}A). Analysis of event occurrence rates (multi-label annotations) further showed that interruptions and gaps occurred more frequently in-person, whereas videoconference sessions were notably more "uneventful" (Figure \ref{fig:annotation_results}B).

These findings suggest that while interruptions and gaps may be superficially categorized as conversational "turbulence," they represent intrinsic elements of natural interaction. Consequently, although videoconferencing produces a "cleaner" exchange, its lack of these natural dynamics may represent a fundamental shift in communication behavior, potentially serving as a root cause of lower rated enjoyment and the phenomenon of "Zoom fatigue" \cite{fauville2023video, kendrick2023turn, bailenson2021nonverbal}.

\section{Conclusion \& Limitations}

Our exploratory analyses highlight the need for having a paired conversation corpus, as the conversational behaviors (both multimodal features and annotations) vary substantially across settings. Specifically, videoconferencing elicits shorter, syntactically simpler speech with longer pauses. In terms of facial expressions, videoconferencing is largely dominated by less salient expressions, punctuated by increases in specific localized expressions. The human annotations further revealed that while in-person conversations feature more interruptions and gaps, they received higher enjoyment ratings than videoconferencing, suggesting that such conversational "turbulence" may be a natural and essential component of a pleasant interaction. Together, such multifaceted differences underscore the complexity of modeling human communication across settings, cautioning against the assumption that models developed in one domain can seamlessly generalize to another. 

Ultimately, VIP-MINGLE aims to facilitate the development of domain-aware models, the refinement of future videoconferencing systems, and a deeper scientific understanding of human communication across both physical and videoconferencing environments. While we utilized a widely adopted semi-structured task for eliciting group dialogue (e.g., \cite{reverdy-etal-2022-roomreader, koutsombogera2018modeling}), we acknowledge that this specific task may not fully generalize to other conversational formats. However, VIP-MINGLE does not aim to maximize task diversity; rather, it serves as a controlled, horizontal bridge to connect research previously isolated within each setting, enabling broader generalization by aligning this corpus with other tasks in each setting in the future.

\section{Generative AI Use Disclosure}
Generative AI tools were used to assist with data analysis code and manuscript editing. The authors remain fully responsible for the accuracy and integrity of the final content.

\section{Acknowledgments}

A.C., D.P., and D.F. are supported by NYU Discovery Research Fund for Human Health. A.C. is supported by Leon Levy Scholarships in Neuroscience, Leon Levy Foundation and New York Academy of Sciences. The funders have no role in study design, data collection and analysis, decision to publish, or preparation of the manuscript. This work was supported in part through the NYU IT High Performance Computing resources, services, and staff expertise.

\bibliographystyle{IEEEtran}
\bibliography{mybib}

\end{document}